\documentclass{article}
\usepackage{spconf,amsmath,graphicx}
\usepackage{amssymb}
\usepackage{mathrsfs}
\usepackage{amsmath}
\usepackage{comment} 
\usepackage{here}
\usepackage{bm}
\usepackage{url}
\usepackage{cite}
\makeatletter
\def\Hline{
  \noalign{\ifnum0=`}\fi\hrule \@height 2.\arrayrulewidth \futurelet
  \reserved@a\@xhline}
\makeatother

\title{
Sound Event Localization based on Sound~Intensity~Vector Refined By DNN-Based Denoising and Source Separation}
\name{Masahiro Yasuda${}^{\dagger}$, Yuma Koizumi${}^{\dagger}$, Shoichiro Saito${}^{\dagger}$, Hisashi Uematsu${}^{\dagger}$ and Keisuke Imoto${}^{\ddagger}$}
\address{
${}^{\dagger}$NTT Media Intelligence Laboratories, Tokyo, Japan\\
${}^{\ddagger}$Ritsumeikan University, Shiga, Japan
}
\begin{document}
\ninept
\maketitle
\begin{abstract}
We propose a direction-of-arrival (DOA) estimation method for Sound Event Localization and Detection (SELD).
Direct estimation of DOA using a deep neural network (DNN), i.e. completely-data-driven approach, achieves high accuracy.
However, there is a gap in the accuracy between DOA estimation for single and overlapping sources because they cannot incorporate physical knowledge.
Meanwhile, although the accuracy of physics-based approaches is inferior to DNN-based approaches, it is robust for overlapping-source.
In this study, we consider a combination of physics-based and DNN-based approaches; the sound intensity vectors (IVs) for physics-based DOA estimation is refined based on DNN-based denoising and source separation.
This method enables the accurate DOA estimation for both single and overlapping sources using 
a
spherical microphone array.
Experimental results show that the proposed method achieves state-of-the-art DOA estimation accuracy on an open dataset of the SELD.
\end{abstract}
\section{Introduction}
\label{sec:intro}
Sound Event Localization and Detection (SELD) is the combined task of Sound Event Localization (SEL) and Sound Event Detection (SED)~\cite{SELD}. 
SEL is the task that identifies when and where a sound event occurred via direction-of-arrival (DOA) and number-of-active-sources (NOAS) estimation. 
SED is the task that classifies each sound event class.
SELD is a fundamental-component for understanding the surrounding environment, and applicable in many applications such as autonomous driving cars~\cite{SmartCar1,SmartCar2} and security systems using drones~\cite{drone}.

The goal with DOA estimation for SELD is to identify the relative position of the sound sources with respect to the microphone at every time frame~\cite{SELD,CNN-DOA}.
Although this task is a physical quantity estimation obviously, most approaches adopt a deep neural network (DNN)--based data-driven approach~\cite{SELD,IV_DNN_DOA,DNN-DOA,CNN-DOA,Kapka_SELD,chang_SELD,Cao_SELD}; using DNN as a regression function for estimating azimuth and elevation directly from observations.
This approach has achieved high-accuracy thanks to the high regression capability of DNN, however, DOA estimation for overlapping sound sources is still a difficult task for a perfectly data-driven approach~\cite{CNN-DOA,Kapka_SELD,IV_DNN_DOA}.
Meanwhile, although the DOA estimation accuracy of the physics-based approach is inferior to that of a single source DNN, it has the advantage of robustness against overlapping sources~\cite{Tho_SELD,Tho_DOA}.
Thus, we consider that there is room for combining the advantages of physics-based and DNN-based methods.

So far, various physics-based DOA estimation methods have been proposed, such as the multiple-signal-classification (MUSIC)~\cite{MUSIC} and the Sound Intensity Vectors (IVs)--based method~\cite{IntensityVector,IV_DOA,AugumentedIntensityVector,DOA_FOAIV,DOA_FOAIV2}.
The MUSIC method can accurately estimate multiple-DOAs, and the IV-based method has a good time-angular resolution.
These characteristics are important requirements for DOA estimation for SELD, thus physics-based method might be suitable for this task.
However, the performance of both methods is degraded in low signal-to-noise ratio (SNR) condition~\cite{CNN-DOA,AugumentedIntensityVector},
that is why DNN-based outperformed these approaches in task 3 of the IEEE AASP Challenge on Detection and Classification of Acoustic Scenes and Events (DCASE 2019 task 3)~\cite{dcase2019web}.
Since several DNN-based methods are also perform well in terms of source enhancement~\cite{DNNMask1,PIT,DNNMask2}, this disadvantage might be able to overcome by combining a DNN as a denoising module with the physics-based DOA estimation procedure.

\begin{figure}[t]
  \centering
  \centerline{\includegraphics[width=75mm]{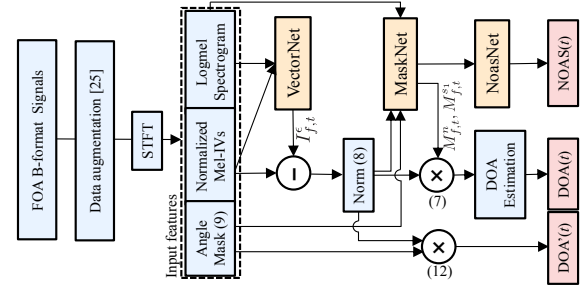}}
  \vspace{-5pt}
  \caption{System overview. 
  Orange, red, and blue boxes denote DNN, output variable, and another variable/operation, respectively.
  IVs obtained from FOA signals is refined by the subtraction of vector $\bm{I}^{\epsilon}_{f,t}$ estimated by VectorNet and the multiplication of two T-F masks $M^{s1}_{f,t},M^n_{f,t}$ estimated by MaskNet.}
  \label{fig:overview}
  \vspace{-2pt}
\end{figure}

We propose a DOA estimation method for overlapping sources by combining IV-based DOA estimation and DNN-based denoising and source separation as shown in Fig.\,\ref{fig:overview}.
IVs obtained from the first-order-ambisonics (FOA) signal is refined using two DNNs, MaskNet and VectorNet, and these DNNs are trained to minimize the error of DOA estimation.
MaskNet estimates two Time-Frequency (T-F) masks for denoising and source separation respectively, and VectorNet removes the components that cannot be removed by the T-F mask due to the large overlap with the target signal at the T-F domain (e.g. reverberation). 
In this study, we assume that the maximum number of overlapping sources are two.

\section{CONVENTIONAL METHODS}
\label{sec:conv}
\subsection{Physics-based methods}
{\bf IV-based method: }
{Ahonen {\it et al.}} proposed a DOA estimation method using IVs calculated from FOA B-format recordings\cite{DOA_FOAIV}. The FOA B-format consists of four channels of signals, and its short-time Fourier transform (STFT) outputs ${\rm W}_{f,t},{\rm X}_{f,t},{\rm Y}_{f,t}$ and ${\rm Z}_{f,t}$ correspond to the 0-th and 1st order of spherical harmonics. Here, $f \in 1,...,F$ and $t \in 1,...,T$ are indexes of frequency and time-frame, respectively. 

In several conventional methods of DOA estimation using IVs, IVs are approximately calculated from the 4-channel spectrograms of the FOA B-format as
\begin{equation}
    {\bf I}_{f,t}\propto \mathfrak{R}\left({\rm W^*}_{f,t}
    \bm{h}_{f,t}
    \right)
    = 
    \left[ 
    I_{X,f,t},
    I_{Y,f,t},
    I_{Z,f,t}
    \right]^{\top},
    \label{eq:int}
\end{equation}
where $\bm{h}_{f,t} = 
\left[ {\rm X}_{f,t}, {\rm Y}_{f,t}, {\rm Z}_{f,t} \right]^{\top}
$,
$\mathfrak{R} (\cdot)$ denotes the real-part of complex numbers, and ${}^*$ is the conjugate of complex numbers.
To select an effective T-F region for DOA estimation, a T-F mask is designed based on the log-power for each bin \cite{DOA_FOAIV}.
Then, the T-F mask is multiplied to IVs, and it is summed on all frequency on each time-frame and obtained time-series IVs.
Finally, DOA of the target source is estimated in each time frame $t$ as
\begin{equation}
\phi_t = \arctan\left(\frac{I_{Y,t}}{I_{X,t}}\right), \;\;\; \theta_t = \arctan\left(\frac{I_{Z,t}}{\sqrt{I_{X,t}^2 + I_{Y,t}^2}}\right),
\label{eq:extract_doa}
\end{equation}
where $\phi \in [-\pi,\pi)$ and $\theta \in [-\pi/2,\pi/2]$ are the azimuth and elevation angle, respectively.
However, since the log-power-based T-F mask cannot separate overlapping sound sources, it is difficult to apply this method for DOA estimation of overlapping sources.

\vspace{5pt}
\noindent
{\bf MUSIC: }
MUSIC~\cite{MUSIC} is a DOA estimation method 
based on the orthogonality between the subspace $E_N$ spanned by the noise steering vector and the steering vector of target sources.
The spatial spectrum $P_{MU}$ derived by MUSIC is defined as follows:
\begin{equation}
P_{MU}(\theta,\phi) = \frac{1}{a^*(\theta,\phi)E_N^*E_Na(\theta,\phi)},
\label{eq:music}
\end{equation}
where $a^*(\theta,\phi)$ is the steering vector in the $(\theta, \phi)$ direction.
If the $a^*(\theta,\phi)$ matches the steering vector of target sources, $P_{MU}$ has a sharp peak due to the orthogonality with $E_N$.
Therefore, the position of this peak corresponds to the DOA of each source.
This method has the advantage that DOA of multiple sound sources can be estimated simultaneously if the number of sound sources is known. However, it is known that this method is not robust against the SNR, as shown in a previous study~\cite{CNN-DOA}.

\subsection{DNN-based methods}
A recent advancement in DOA estimation is the use of DNN as a regression function for directly estimating the azimuth and elevation labels from the observations \cite{IV_DNN_DOA,SELD,DNN-DOA,CNN-DOA,Kapka_SELD,chang_SELD,Cao_SELD}.
Several DNN-based methods outperforms conventional parametric DOA estimation methods without any physical knowledge.
In fact, many participants of DCASE challenge 2019 task 3 used perfectly data-driven approaches for DOA estimation  \cite{Kapka_SELD,chang_SELD,Cao_SELD} and achieved good accuracy.
With these methods, the DNN structure is 
a combination of a multi-layer CNN and bidirectional-gated recurrent units (Bi-GRUs), which enable extraction of higher-order features and modeling of temporal structure,
and the DNN is trained to minimize a metric between the true and estimated DOA labels such as the mean-absolute-error (MAE).
However, DOA estimation of overlapping sources is difficult for such a data-driven DNN-based method, and it is reported that the accuracy is much lower than the case of single source~\cite{IV_DNN_DOA,CNN-DOA,Kapka_SELD}.

\section{PROPOSED METHOD}
\label{sec:proposed}
\subsection{Basic concept}
Our DOA estimation method for overlapping sources is based on IVs refined by DNN-based denoising and source separation.
Generally, a time-domain input signal $\bm{x}$ can be expressed as
\begin{equation}
    \bm{x} = \sum_{i=1}^{N}\bm{s}_i + \bm{n} + \bm{\epsilon},
    \label{eq:x_srn}
\end{equation}
where $\bm{s}_i$ is the direct sound of sound source $i \in \{1,...,N\}$, $\bm{n}$ is noise uncorrelated to the sound source, and $\bm{\epsilon}$ is the other terms related to target sources (e.g. room reverberation).
According to this modeling, the T-F representation $\bm{x}$ can also be written as the sum of the same components.
Thus, time series of IVs calculated by (\ref{eq:int}) can be expressed as
\begin{equation}
    {\bf I}_{t} =  \sum_{f=1}^{F}\left(\sum_{i=1}^{N}{\bf I}^{s_i}_{f,t} + {\bf I}^n_{f,t} + {\bf I}^{\epsilon}_{f,t} \right).
    \label{eq:int3}
\end{equation}
As we can see in (\ref{eq:int3}) the observed IVs ${\bf I}_{t}$ are affected by not only the $i$th sound source but also other
components.
This is one of the reasons for weak points of the conventional IV-based method. 

To overcome this, we refine observed IVs via denoising and source separation by multiplying T-F mask, and suppression of the $\epsilon$-component by vector subtraction.
Usually, T-F mask-based source separation assumes that all components are sufficiently sparse in the T-F doma
in~\cite{BSS}.
In practice, this is a strong assumption, and it is not possible to assume sufficient sparsity especially for noise components.
For this reason, we use the combination of T-F masks $M^{s_i}_{f,t}(1-M^n_{f,t})$ 
where $M^{s_i}_{f,t}$ extracts the $i$-th sound source from $\sum_{i=1}^N {\bf I}^{s_i}_{f,t}$ and $M^n_{f,t}$ extract ${\bf I}^n_{f,t}$ from the observation.
If $\bm{\epsilon}$ includes reverberation components of $\bm{s}_i$, ${\bf I}^{\epsilon}_{f,t}$ could have large overlap with ${\bf I}^{s_i}_{f,t}$ in T-F domain and cannot be removed with a T-F mask. 
Thus, we estimated this term as a vector and subtracted it directly from ${\bf I}_{f,t}$.
This process can be written as
\begin{equation}
{\bf I}^{s_i}_{t} = \sum_{f=1}^{F}M_{f,t}^{s_i}*(1 - M_{f,t}^n)*\left( {\bf I}_{f,t} - \hat{{\bf I}}^{\epsilon}_{f,t} \right).
\label{eq:int_s}
\end{equation}
In this paper, we consider only when the maximum number of sound sources is two. 
At this time, we can use $(1-M^{s_1}_{f,t})$ instead of $M^{s_2}_{f,t}$.
Therefore, what we should estimate are $M^{s_1}_{f,t}, M^{n}_{f,t}$ and $\hat{{\bf I}}^{\epsilon}_{f,t}$, and we estimate them using two DNNs as shown in Fig.\,\ref{fig:overview}.
\subsection{Network architecture}
\begin{figure}[t]
  \centering
  \centerline{\includegraphics[width=80mm]{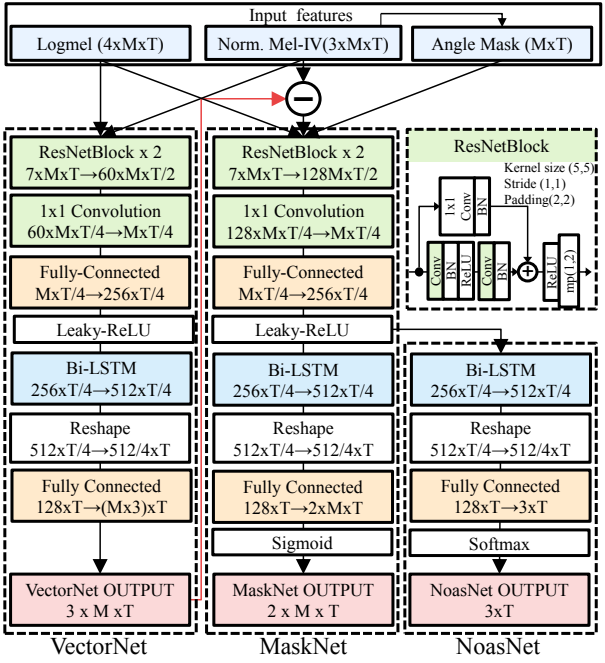}}
  \vspace{-5pt}
  \caption{DNN architecture of the proposed method. In the figure of 
  ResNetBlock part, ``Conv'', ``BN'', and ``Maxpool'' denotes convolutional layer, batch normalization, and max pooling, respectively.}
  \label{fig:DNNmodel}
  \vspace{-2pt}
\end{figure}
\subsubsection{Input features}
As Fig.\,\ref{fig:DNNmodel} shows, there are three input features to DNNs.
One is a logmel-spectrogram of the input signal. 
The second is the sound intensity vector ${\bf I}_{f,t}$ of the input signal.
IVs also are compressed by Mel-filterbank to guarantee the dimension of IVs is the same as that of the logmel-spectrograms like as~\cite{Cao_SELD}. In addition, the IVs are normalized as:
\begin{equation}
    {\bf I}^{\rm norm}_{f,t} = \frac{{\bf I}_{f,t}}{|{\bf I}_{f,t}|},
    \label{eq:norm}
\end{equation}
because only the direction of IVs is necessary for DOA estimation by (\ref{eq:extract_doa}).
The third is the angle mask $M^{angle}_{f,t}$, which is an elementary T-F mask for source separation derived directly from observed IVs. 
This T-F mask is defined as
\begin{equation}
M^{angle}_{f,t} = {\rm sigmoid}\left(\angle\left({\bf I}_{XY,f,t}{\bf R}\left(-\phi^{a}_t\right)\right)\right),
\label{eq:mangle}
\end{equation}
where $\angle{\bf I}_{XY}=\arctan(I_Y/I_X)$, ${\bf R}(\phi)$ is the rotation matrix for azimuth $\phi$, and $\phi^a_t$ is the azimuth angle of 
${\bf I}^{a}_t=\sum_f{\bf I}^{\rm norm}_{f,t}$.
This mask passes $(f,t)$-th T-F element where ${\bf I}_{f,t}$ points in the counterclockwise direction than the reference direction $\phi^a_t$.
When NOAS equals to two, the reference direction of ${\bf I}^a_t$ indicates a direction between two sources' DOAs, and either one of the sources exists in the counterclockwise direction.
Therefore, by rotating the ${\bf I}_{f,t}$ of such a source by $-\phi^a_t$, the sigmoid argument becomes positive.
We used this mask as the DNN input and also used it in the regularization term of the loss function (see Sec\,\ref{sec:loss}).

\subsubsection{DNN model architecture}
Figure\,\ref{fig:DNNmodel} shows the proposed DNN architecture. The DNN model consists of three parts.
The first DNN, called VectorNet, takes logmel-spectrogram and normalized Mel-IVs as inputs and estimates ${\bf I}^{\epsilon}_{f,t}$ term in (\ref{eq:int_s}). Next, refined IVs ${\bf I}'_{f,t}={\bf I}_{f,t}-\hat{{\bf I}}^{\epsilon}_{f,t}$, logmel-spectrograms and the angle mask $ M^{angle}_{f,t}$ are inputted to MaskNet. Note that ${\bf I}'_{f,t}$ is also normalized by (\ref{eq:norm}).
In MaskNet, the denoising mask $1-M^n_{f,t}$ and the source separation mask $M^{s_1}_{f,t}$ are estimated and output as the concatenated form.
Finally, NOAS is estimated using the DNN that is a branch of MaskNet with softmax activation called NoasNet.
VectorNet and MaskNet are composed of multi-layer CNN blocks for high-level feature extraction and RNN layers for temporal structure modeling.
Final estimates of azimuth and elevation are calculated by (\ref{eq:extract_doa}) from IVs refined by (\ref{eq:int_s}) using VectorNet and MaskNet outputs. Note that if the estimated NOAS=1, then $M^{s_1}_{f,t}=1 $ is used.

\begin{figure*}[t]
\centering
  \centerline{\includegraphics[angle=90,width=175mm]{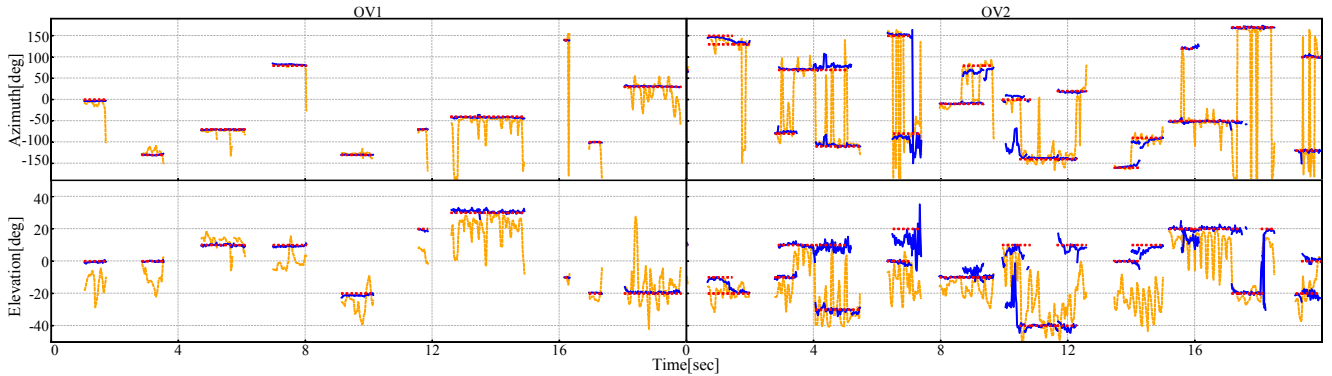}}
  \vspace{-5pt}
  \caption{Example of DOA estimation result. 
  Red-dotted line shows ground truth, orange-dashed line shows DOAs estimated by original-IVs and blue-solid line shows DOAs estimated by (B) without post-processing (\ref{eq:pps}).}
  \label{fig:DOAresult}
  \vspace{-10pt}
\end{figure*}
\subsection{Loss Function}
\label{sec:loss}
As the loss function, we used the mean-absolute-error (MAE) loss $\mathcal{L}^{DOA}$ for the DOA estimation, binary-cross-entropy (BCE) loss $\mathcal{L}^{NOAS}$ for the NOAS estimaion with the one-hot label. To train the DNNs, we used the sum of these loss functions and simultaneously trained all networks in an end-to-end manner. 

Since DOA is a phase variable, the difference in the estimate and label of source directions must be less than $\pi$. To guarantee this, we define the rotational-MAE loss as:
\begin{equation}
    \begin{split}
        \Delta\theta_t &= |\hat{\theta}_t - \theta_t|,\\
        \Delta\phi_t &= {\rm min}\left(|\hat{\phi}_t - \phi_t|,|\hat{\phi}_t\pm{2\pi} - \phi_t|\right), 
    \end{split}
    \label{eq:rotmae}
\end{equation}
respectively. 
Besides, considering NOAS = 2, we cannot decide which ground truth DOA $(\theta^j,\phi^j)_{j\in{1,2}}$ is correspond to the predicted DOA $(\hat{\theta}^i,\hat{\phi}^i)_{i\in{1,2}}$. Because of this permutation problem, we used the following loss function:
\begin{equation}
    \mathcal{L}^{DOA} = \frac{1}{\mathcal{Z}}\sum_{t=1}^{T} z_t {\rm min}
    (\Delta D^{11} + \Delta D^{22},\Delta D^{12} + \Delta D^{21}),
    \label{eq:ldoa}
\end{equation}
where $D^{ij}$ is the rotational-MAE loss between $(\hat{\theta}^i,\hat{\phi}^i)$ and $(\theta^j,\phi^j)$, $z_t$ is the ground truth of NOAS, and normalization term $\mathcal{Z}$ is defined as $\mathcal{Z}= \sum_{t=1}^T z_t$. 
In addition, as we can see from Fig.\,\ref{fig:overview}, $\mathcal{L}^{DOA}$ is far from the VectorNet output, and it may result in the gradient vanishing problem.
To avoid this, we additionally use 
the MaskNet independent DOA loss $\mathcal{L}^{DOA'}$
as the regularization term. The DOA$'$ $(\theta'^i_t, \phi'^i_t)$ is derived using MaskNet independent refined-IVs that defined as:
\begin{equation}
    {\bf I'}_{f,t} = \sum_{f=1}^{F}M_{f,t}^{angle}*({\bf I}_{f,t} - \hat{{\bf I}}^{\epsilon}_{f,t}).
\end{equation}
By using this DOA$'$, $\mathcal{L}^{DOA'}$ is calculated by using (\ref{eq:rotmae}), (\ref{eq:ldoa}).
Therefore, the overall loss function is thus expressed as:
\begin{align}
    \mathcal{L} &=\mathcal{L}^{DOA} + \lambda_1\mathcal{L}^{NOAS} + \lambda_2\mathcal{L}^{DOA'},
\end{align}
where $\lambda_{1,2}$ are hyperparameters.

\section{EXPERIMENTS}
\subsection{Experimental Setup}
The experiment was performed using 500 FOA recordings on the TAU Spatial Sound Events 2019 dataset~\cite{DCASE_dataset}. 500 sound recordings consisted of 0 to 4 splits.
Each split contains the same number of subsets with the maximum NOAS equals to 1 and 2 (OV1 and 2).
We fold the 5 splits into two type of the experimental set.
One is the ``MainSet'' that has 400 training data and 100 test data, second is the ``DevSet'' that has 4 combination of 200 training data, 100 validation data and 100 test data. To augment the training data, we used the \textit{FOA-domain spatial Augmentation}~\cite{FOA_aug}.

In all experiments, the sampling frequency was 48 kHz. 
For the STFT, a 8192-point Hanning window with 960-point shift was utilized.
The number of mel-filter-banks applied to the spectrogram and the IV was set to 96. 
We used ADAM~\cite{Adam} optimizer with initial learning rate $\alpha=0.001$.
This learning rate was decreased step-wize by a factor of 0.1 at the 150, 225 epoch.
We always conclude training after 300 epochs.
The loss weight $\lambda_{1,2}$ was set to 10, 0.1 respectively.

The estimated DOA and NOAS are post-processed by the following procedure. First, estimated NOAS is obtained from argmax of the NoasNet output. Next, after smoothing the Noas output using the same technique as Kapka {\it et al.}~\cite{Kapka_SE
LD}, we determined the onset and offset for each event.
Here, an event is defined as an interval where NOAS is constant.
In addition, since test datasets is known to have sound sources in 10$^\circ$ steps, the obtained DOAs are discretized at 10$^\circ$ intervals. Furthermore, for smoothing, the median value for DOA in the event is taken as the DOA of that event:
\begin{equation}
    \begin{split}
        {\rm DOA_{dis}} &= {\rm round}({\rm DOA}/10^\circ)*10^\circ \\
        {\rm DOA_{med}} &= {\rm median}({\rm DOA_{dis}}[{\rm onset\ time:offset\ time}]).
    \end{split}
    \label{eq:pps}
\end{equation}

For a fair comparison with previous research on SELD, we perform SED experiments using the same method with Kapka {\it et al.}~\cite{Kapka_SELD}, because their method can perform SED independently of DOA estimation. SED is inferred by the DNN model that has combination of CNN and RNN architecture with 4CH log-spectrogram input and then combined with DOA estimation results. Although the correspondence between DOA and SED can not determined uniquely when NOAS equals to 2, it is estimated using DOA and SED of the preceding and following time-frames that have NOAS equals to 1 as post-processing.

\subsection{Result}
\begin{table}[]
  \centering
  \caption{Experimental results using ``Mainset''. DE, FR, ER and F denotes DOA-error, Frame-recall, error rate and F-score.} 
\begin{tabular*}{88mm}{l|cc|cc}
\Hline
                                   & DE     & FR        & ER       & F \\ \hline
Nguyen {\it et al.}\cite{Tho_SELD} (Phys.-based) & 5.4$^{\circ}$    & 0.888     & 0.11     & 0.934 \\
Kapka {\it et al.}\cite{Kapka_SELD} (DNN-based)& 3.7$^{\circ}$    & {\bf0.968}& {\bf0.08}& {\bf0.947} \\ 
Noh {\it et al.}\cite{chang_SELD} (DNN-based)& 2.7$^{\circ}$    & 0.908     & 0.14     & 0.919 \\ \hline
(A): Mask (Combination)                         & 8.3$^{\circ}$      &0.910     & 0.17        & 0.888 \\
(B): Mask+Vector (Combination)&{\bf2.2$^{\circ}$}& 0.956     & 0.12     & 0.909 \\ 
\Hline
\end{tabular*}
  \label{tb:result}
  \vspace{-5pt}
\end{table}

Evaluation was performed using DOA-error (DE), Frame-recall (FR), error rate (ER) and F-score (F) as metrics, which were used in DCASE2019 Challenge - task3~(cf.~\cite{Metrics}).
DE represents the error of the estimated angle, and FR represents the recall of NOAS estimation. These are metrics related to SEL.
On the other hand, ER and F are metrics related to SED, where ER is the amount of errors and F is the harmonic average of accuracy and recall. In order to step-wisely confirm the effectiveness of the VectorNet and MaskNet, we tested 2 patterns of the proposed method: (A) does not have VectorNet, (B) full-architecture.
Proposed method was compared with two DNN-based methods~\cite{Kapka_SELD,chang_SELD} and one physics-based method~\cite{Tho_SELD} that were evaluated using the same dataset and metrics.
In DCASE2019 challenge task3, \cite{Kapka_SELD}, \cite{chang_SELD} and \cite{Tho_SELD} achieved the best overall score, the best DE score and the best DE score as physics-based method, respectively. 
Both DNN-based methods take a perfectly data-driven approach.
The physics-based method uses the eigenvectors of the spatial correlation matrix as in MUSIC. Table 1 shows the experimental results of SELD task. Results shows that DE of (B) is lower than (A), indicating that the combination of VectorNet and MaskNet is effective in improving IVs.  Furthermore, the DE of (B) is always lower than conventional methods, achieving state-of-the-art accuracy.
Table 2 shows the DE and FR results for the OV1,2 subset. Compared to the DNN-based conventional method, DE of (B) is improved especially in the case of OV2, and it also comparable with physics-based method. This result shows the effectiveness of the source separation using T-F mask estimated by MaskNet.
Fig.\,\ref{fig:DOAresult} shows an example of DOA estimation using model-B without post-processing (\ref{eq:pps}), and we can be seen that original-IVs is refined by applying (\ref{eq:int_s})  and approaches the ground truth.
Therefore, we conclude that the accuracy of parametric-based DOA estimation is improved by refinement of physical parameters using DNN-based denoising and source separation.

\begin{table}[]
\centering
\caption{Experimental results using OV1,2 subsets of ``Devset''.}
\begin{tabular}{l|l|cc}
\Hline
    &                                       & DE & FR \\ \hline
     & Nguyen {\it et al.}\cite{Tho_SELD} (Phys.-based)  & 4.7$^{\circ}$ & 0.96     \\ 
OV1  & Kapka {\it et al.}\cite{Kapka_SELD} (DNN-based) & 1.3$^{\circ}$ & {\bf 0.99} \\ \cline{2-4}
     & (B): Mask+Vector (Combination) & {\bf 1.1$^{\circ}$} & 0.98         \\\Hline
     & Nguyen {\it et al.}\cite{Tho_SELD} (Phys.-based)  & {\bf 5.4$^{\circ}$} & 0.82    \\ 
OV2  & Kapka {\it et al.} \cite{Kapka_SELD} (DNN-based) & 7.9$^{\circ}$& {\bf 0.93}  \\ \cline{2-4}
     & (B): Mask+Vector (Combination) & 5.6$^{\circ}$& 0.90         \\\Hline
\end{tabular}
\label{tb:result2}
  \vspace{-5pt}
\end{table}

\section{CONCLUSION}
We proposed a  method for refining IV-based DOA estimation via DNN-based denoising and source separation.
This refinement is done by multiplying the T-F mask for denoising and source separation, and subtracting noise components that cannot be removed by the T-F mask.
Through objective experiments on a  DOA estimation of overlapping sources, we confirmed that the proposed method outperformed a conventional IV-based and DNN-based DOA estimation methods, and the average DE of the proposed method was 2.2$^{\circ}$. 
We conclude that DNN-based denoising and source separation are effective in improving IV-based DOA estimation.
\label{sec:cncl}

\newpage


\begin{thebibliography}{10}

\bibitem{SELD}
Sharath Adavanne, Archontis Politis, Joonas Nikunen, and Tuomas Virtanen.
\newblock Sound event localization and detection of overlapping sources using
  convolutional recurrent neural networks.
\newblock {\em IEEE Journal of selected topics in signal processing}, 13, 2019.

\bibitem{SmartCar1}
Yong Xu, Qiuqiang Kong, Wenwu Wang, and Mark~D. Plumbley.
\newblock Surrey-cvssp system for {DCASE}2017 challenge task4.
\newblock In {\em Tech. report of Detection and Classification of Acoustic
  Scenes and Events 2017 ({DCASE}) Challange}, 2017.

\bibitem{SmartCar2}
Donmoon Lee, Subin Lee, Yoonchang Han, and Kyogu Lee.
\newblock Ensemble of convolutional neural networks for weakly-supervised sound
  event detection using multiple scale input.
\newblock In {\em Tech. report of Detection and Classification of Acoustic
  Scenes and Events 2017 ({DCASE}) Challange}, 2017.

\bibitem{drone}
Xianyu Chang, Chaoqun Yang, Xiufang Shi, Pengfei Li, Zhiguo Shi, and Jiming
  Chen.
\newblock Feature extracted {DOA} estimation algorithm using acoustic array for
  drone surveillance.
\newblock In {\em Proc. of IEEE 87th Vehicular Technology Conference}, 2018.

\bibitem{CNN-DOA}
Sharath Adavanne, Archontis Politis, and Tuomas Virtanen.
\newblock Direction of arrival estimation for multiple sound sources using
  convolutional recurrent neural network.
\newblock In {\em Proc. of IEEE 26th European Signal Processing Conference},
  2018.

\bibitem{IV_DNN_DOA}
Laur\'{e}line Perotin, Romain Serizel, Emmanuel Vincent, and Alexandre Guerin.
\newblock {CRNN}-based multiple {D}o{A} estimation using acoustic intensity
  features for ambisonics recordings.
\newblock {\em IEEE Trans. Signal Process.}, 13(1), 2019.

\bibitem{DNN-DOA}
Zhang-Meng Liu, Chenwei Zhang, and Philip~S. Yu.
\newblock Direction-of-arrival estimation based on deep neural networks with
  robustness to array imperfections.
\newblock {\em IEEE Transactions on Antennas and Propagation}, 66:7315--7327,
  2018.

\bibitem{Kapka_SELD}
S$\l$awomir Kapka and Mateusz Lewandowski.
\newblock Sound source detection, localization and classification using
  consecutive ensemble of {CRNN} models.
\newblock In {\em Tech. report of Detection and Classification of Acoustic
  Scenes and Events 2019 ({DCASE}) Challange}, 2019.

\bibitem{chang_SELD}
Kyoungjin Noh, Jeong-Hwan Choi, Dongyeoup Jeon, and Joon-Hyuk Chang.
\newblock Three-stage approach for sound event localization and detection.
\newblock In {\em Tech. report of Detection and Classification of Acoustic
  Scenes and Events 2019 ({DCASE}) Challange}, 2019.

\bibitem{Cao_SELD}
Yin Cao, Turab Iqbal, Qiuqiang Kong, Miguel~B. Galindo, Wenwu Wang, and Mark~D.
  Plumbley.
\newblock Two-stage sound event localization and detection using intensity
  vector and generalized cross-correlation.
\newblock In {\em Tech. report of Detection and Classification of Acoustic
  Scenes and Events 2019 ({DCASE}) Challange}, 2019.

\bibitem{Tho_SELD}
Thi Ngoc~Tho Nguyen, Douglas~L. Jones, Rishabh Ranjan, Sathish Jayabalan, and
  Woon~Seng Gan.
\newblock {DCASE} 2019 task 3: A two-step system for sound event localization
  and detection.
\newblock In {\em Tech. report of Detection and Classification of Acoustic
  Scenes and Events 2019 ({DCASE}) Challange}, 2019.

\bibitem{Tho_DOA}
Nguyen Thi~Ngoc Tho, Shengkui Zhao, and Douglas~L Jones.
\newblock Robust {DOA} estimation of multiple speech sources.
\newblock In {\em Proc. of IEEE International Conference on Acoustics, Speech
  and Signal Processing (ICASSP)}, 2014.

\bibitem{MUSIC}
Ralph~O. Schmidt.
\newblock Multiple emitter location and signal parameter estimation.
\newblock {\em IEEE Transactions On Antennas and propagation}, 34:276--280,
  1986.

\bibitem{IntensityVector}
D.~P. Jarrett, E.~A.~P. Habets, and P.~A. Naylor.
\newblock {3D} source localization in the spherical harmonic domain using a
  pseudointensity vector.
\newblock In {\em Proc. of 18th European Signal Processing Conference}, 2010.

\bibitem{IV_DOA}
Despoina Pavlidi, Symeon Delikaris-Manias, Ville Pulkki, and Athanasios
  Mouchtaris.
\newblock {3D} localization of multiple sound sources with intensity vector
  estimates in single source zones.
\newblock In {\em Proc. of 23rd European Signal Processing Conference}, 2015.

\bibitem{AugumentedIntensityVector}
Sina Hafezi, Alastair~H. Moore, and Patrick~A. Naylor.
\newblock Augmented intensity vectors for direction of arrival estimation in
  the spherical harmonic domain.
\newblock volume~25, pages 1956--1968, 2017.

\bibitem{DOA_FOAIV}
J.~Ahonen, V.~Pulkki, and T.~Lokki.
\newblock Teleconference application and {B}-format microphone array for
  directional audio coding.
\newblock In {\em Proc. of AES 30th International Conference: Intelligent Audio
  Environments}, 2007.

\bibitem{DOA_FOAIV2}
Srdan Kiti\'c and Alexandre Gu\'erin.
\newblock {TRAMP}: Tracking by a real-time ambisonic-based particle filter.
\newblock In {\em Proc. of LOCATA ChallengeWorkshop, a satellite event of
  IWAENC}, 2018.

\bibitem{dcase2019web}
{{DCASE}2019 Workshop}.
\newblock \url{http://dcase.community/workshop2019/}.

\bibitem{DNNMask1}
Hakan Erdogan, John~R. Hershey, Shinji Watanabe, and Jonathan~Le Roux.
\newblock Phase-sensitive and recognition-boosted speech separation using deep
  recurrent neural networks.
\newblock In {\em Proc. of IEEE International Conference on Acoustics, Speech
  and Signal Processing (ICASSP)}, 2015.

\bibitem{PIT}
Morten Kolb$\ae$k, Dong Yu, Zheng~Hua Tan, and Jesper Jensen.
\newblock Multitalker speech separation with utterance-level permutation
  invariant training of deep recurrent neural networks.
\newblock {\em IEEE/ACM Transactions on Audio, Speech, and Language
  Processing}, 25:1901--1913, 2017.

\bibitem{DNNMask2}
Yuma Koizumi, Kenta Niwa, Yusuke Hioka, Kazunori Kobayashi, and Yoichi Haneda.
\newblock {DNN}-based source enhancement to increase objective sound quality
  assessment score.
\newblock {\em IEEE ACM Transactions on Audio, Speech, and Language
  Processing}, 26:1780--1792, 2018.

\bibitem{BSS}
Ozgur Y脹lmaz and Scott Rickard.
\newblock Blind separation of speech mixtures via time-frequency masking.
\newblock {\em IEEE Trans. Signal Process.}, 52:1830--1847, July. 2004.

\bibitem{DCASE_dataset}
Sharath Adavanne, Archontis Politis, and Tuomas Virtanen.
\newblock A multi-room reverberant dataset for sound event localization and
  detection.
\newblock {\em arXiv1905.08546v2}, 2019.

\bibitem{FOA_aug}
Luca Mazzon, Yuma Koizumi, Masahiro Yasuda, and Noboru Harada.
\newblock First order ambisonics domain spatial augmentation for {DNN}-based
  direction of arrival estimation.
\newblock In {\em Proc. of 4th Workshop on Detection and Classification of
  Acoustic Scenes and Events 2019 ({DCASE})}, 2019.

\bibitem{Adam}
Diederik~P. Kingma and Jimmy~Lei Ba.
\newblock Adam: A method for stochastic optimization.
\newblock {\em arXiv:1412.6980v9}, 2017.

\bibitem{Metrics}
Annamaria Mesaros, Sharath Adavanne, Archontis Politis, Toni Heittola, and
  Tuomas Virtanen.
\newblock Joint measurement of localization and detection of sound events.
\newblock In {\em Proc. of IEEE Workshop on Applications of Signal Processing
  to Audio and Acoustics (WASPAA)}, 2019.

\end{thebibliography}
\end{document}